\documentclass{PoS}

\usepackage[english]{babel} 

\usepackage{tabularx}
\usepackage[singlelinecheck=false]{caption}
\usepackage{booktabs}
\usepackage[caption=false]{subfig}
\usepackage[T1]{fontenc}
\usepackage{soul}
\usepackage{enumitem}
\usepackage{xcolor}
\usepackage[titletoc]{appendix}
\usepackage{overpic}
\usepackage[normalem]{ulem}
\usepackage[nointegrals]{wasysym}
\usepackage{tablefootnote}
\usepackage[caption=false]{subfig}
\usepackage{graphicx}
\usepackage{nicefrac}
\usepackage{tabularx}
\usepackage{booktabs}
\usepackage{multirow}
\usepackage{amssymb,amsmath}
\usepackage{footnote}
\usepackage{bm}
\usepackage[capitalize]{cleveref}[2012/02/15]

\usepackage{lineno}
\usepackage{accents}

\usepackage{lineno}

\usepackage{tabu}
\usepackage{url}
\usepackage{siunitx}
\DeclareSIUnit\parsec{pc}
\DeclareSIUnit\lightyear{ly}

\let\OldAng\ang%
\renewcommand*{\ang}[2][]{%
    \OldAng[scientific-notation=false,separate-uncertainty=true,round-mode=places,round-precision=2,#1]{#2}%
}
\DeclareSIUnit\year{yr}
\DeclareSIUnit\erg{erg}
\DeclareSIUnit\msun{M_{\astrosun}}
\DeclareSIUnit{\GeV}{\giga\electronvolt}
\DeclareSIUnit{\TeV}{\tera\electronvolt}
\DeclareSIUnit{\PeV}{\peta\electronvolt}
\DeclareSIUnit{\MeV}{\mega\electronvolt}
\DeclareSIUnit{\eV}{\electronvolt}
\DeclareSIUnit{\smm}{\square\metre\second}
\DeclareSIUnit{\smmr}{\metre^{-2}\second^{-1}}
\DeclareSIUnit{\dc}{d.c.}
\DeclareSIUnit{\pe}{p.e.}
\DeclareSIUnit{\nucleon}{nucleon}

\definecolor{desyOrange}{RGB}{242,142,0}
\usepackage{csquotes}
\usepackage[style=authoryear,giveninits=true,%
natbib=true,backend=bibtex,maxcitenames=2, style=numeric,
    sortcites,
    sorting=none,
doi=false,isbn=false,url=false,eprint=false]{biblatex}
\DeclareNameAlias{sortname}{last-first}

\usepackage{footnote}
\makesavenoteenv{tabular}
\makesavenoteenv{table}

\title{A Survey of TeV emission from Galactic Supernova Remnants with HAWC }

\ShortTitle{A Survey of TeV emission from Galactic Supernova Remnants with HAWC}

\author{\speaker{Henrike Fleischhack}\\
        Michigan Technological University\\
        E-mail: \email{hfleisch@mtu.edu}}

\author{for the HAWC collaboration \thanks{Full author list and acknowledgements:  PoS(ICRC2019)1177 and \protect\url{http://www.hawc-observatory.org/collaboration/icrc2019.php}}}

\abstract{Supernova remnants (SNRs) have long been hypothesized as the main source of Galactic Cosmic Rays up to PeV energies. Some of them have indeed been shown to accelerate protons to TeV energies and above. But which of them are indeed efficient accelerators of protons and nuclei? And up to which energies can they accelerate these particles? Measurements of non-thermal emission, especially in the X-ray and gamma-ray regimes, are essential to answer these questions.\\

The High-Altitude Water Cherenkov (HAWC) observatory, surveying the northern TeV gamma-ray sky, is currently the most sensitive wide field-of-view survey instrument in the VHE (very-high-energy, $>$100 GeV) range. With more than three years of data recorded, HAWC is ideally suited for an unbiased survey of gamma-ray emission from galactic SNRs, particularly at TeV energies and above. In this proceeding, I will give an overview of recent measurements of VHE gamma-ray emission from SNRs with the HAWC Observatory. Combined with data from other wavelengths, these measurements are used to derive information about the underlying particle populations such as the maximum acceleration energy and whether leptonic or hadronic processes are responsible for the emission. }

\FullConference{36th International Cosmic Ray Conference -ICRC2019-\\
		July 24th - August 1st, 2019\\
		Madison, WI, U.S.A.}

\begin{document}

\section{Introduction}
At the end of the lifetime of a massive star, its core will collapse to a neutron star or black hole, while its outer shell explodes outward, carrying on the order of $10^{51}$\,\si{\erg} of kinetic energy. The resulting shockwaves can continue to expand into the surrounding medium for tens of thousands of years or more. 

These shocks can accelerate particles (electrons and protons/nuclei from the ISM and the SNR ejecta) to relativistic energies. Given galactic supernova rates and their average energy release, supernova remants are thought to be the main source of Galactic cosmic rays. Indeed, two SNRs have been shown to accelerate protons to relativistic energies (pion bump). But are SNRs really able to drive cosmic ray acceleration up to the purported end of the Galactic cosmic ray spectrum, the knee at a few PeVs? Gamma rays, in particular TeV gamma-rays, are tracers for high-energy particles, and allow us to study up to which energies supernova remnants can  accelerate particles.

In 2016, the Fermi-LAT collaboration conducted a search for gamma-ray emission fom radio-detected galactic SNRs in the \SIrange{1}{100}{\GeV} range, using 3 years of Fermi-LAT data \citep{Acero_2016}. They found thirty gamma-ray sources likely associated with galactic SNRs, and fourteen sources 'marginally classified' as SNRs. In this sudy, we search for TeV gamma-ray emission from a subsample of these sources.

\section{The HAWC Detector and Data Analysis}
The HAWC (High Altitude Water Cherenkov) Observatory is a gamma-ray observatory located at an altitude of 4100 m near volcano Sierra Negra in the state of Puebla, Mexico. Its 1200 photomultiplier tubes in 300 water tanks detect Cherenkov emission from the electromagnectic component of gamma-ray and cosmic-ray induced air showers. More information about the HAWC detector, event reconstruction, and data analysis can be found in \cite{2HWC, Abeysekara:2017mjj}.

HAWC's angular resolution depends on the shower energy, its zenith angle, and the position of the shower core relative to the center of the array. It varies between roughly \ang{0.1} and \ang{1.0}. The energy range over which HAWC is sensitive is different for each source and depends on the source's spectrum as well as its declination (sources with a declination of +\ang{19} will transit directly overhead where HAWC is most sensitive; sources with a higher or lower declination will have a smaller maximum elevation). HAWC's energy threshold ranges from about \SI{500}{\GeV} (soft-spectrum sources at favorable declinations) to \SI{20}{\TeV} (hard-spectrum sources at unfavorable declinations) \cite{2HWC}. Here, the energy range is defined as the central interval contributing 75\% of the test statistic of a source with a given spectrum and at a given declination.

Two data samples were used for this study. For the upper-limit calculation, 1128 days of HAWC data were used, reconstructed and binned according to the 9-bin scheme described in \cite{Abeysekara:2017mjj}. For spectral fits, a smaller dataset (1038 days) of HAWC data was used, which was reconstructed with the ground parameter energy estimator and binned according to the two-dimensional scheme described in \cite{Abeysekara:2019edl}. 

HAWC uses a forward-folding likelihood analysis. More details can be found in \cite{2015arXiv150807479Y}. For this study, SNR morphologies were modeled either as flat, radially symmetric disks or as point sources. SNR energy spectra were modeled as power laws,

\begin{equation}
\frac{dN}{dE} = K \cdot \left( \frac{E}{E_0} \right) ^{-\gamma},
\end{equation}

with flux normalization $K$ at pivot energy $E_0$ and spectral index $\gamma$.

To assess whether an excess of gamma-ray events above background expectations is statistically significant, the following test statistic (TS) is used:

\begin{equation}
TS = -2 \frac{\ln \mathcal{L}}{\ln \mathcal{L}_0},
\end{equation}
where $\mathcal{L}$ is the best-fit likelihood including the source in question, and $\mathcal{L}_0$ is the likelihood of a model without that source.Generally, sources with $TS>25$ are classified as 'detected'.

\subsection{SNR Sample}

\begin{table}[p]
\caption{GeV-detected SNRs in HAWC's field of view. RA and Dec refer to the centroid of the GeV emission from \citep{Acero_2016}. TeV association infomation from TeVCat (\protect\url{www.tevcat.org}). \\ *Notes: SNR205.5+00.5 overlaps with HAWC J0635+070. SNR078.2+02.1 overlaps with Cygnus Cocoon, see \cite{gCygniProceedings, cocoon} for detailed multi-source studies.}
\begin{tabular*}{\textwidth}{@{\extracolsep{\fill} } lccccc}
\toprule
\textbf{SNR name} & \textbf{RA} & \textbf{Dec} & \textbf{TeV association} & \textbf{Isolated} & \textbf{HAWC} \\
&  &  &  &  & \textbf{detected} \\
\midrule
SNR006.4-00.1 & \ang{ 270.36 } &  \ang{ -23.44} & HESS J1801-233 \cite{W29}  & no & --- \\ 
SNR008.7-00.1 & \ang{ 271.36 } &  \ang{ -21.59} &  HESS J1804-216 \cite{Aharonian_2006} &  no & --- \\ 
SNR020.0-00.2 & \ang{ 277.11 } &  \ang{ -11.50} & ---  & no & --- \\ 
SNR023.3-00.3 & \ang{ 278.57 } &  \ang{ -8.75} & HESS J1834-087 \cite{Aharonian_2006,Albert_2006}  & no & --- \\ 
SNR024.7+00.6 & \ang{ 278.60 } &  \ang{ -7.17} &  --- & no & --- \\ 
SNR034.7-00.4 & \ang{ 284.05 } &  \ang{ 1.34} &  ---  & no & --- \\ 
SNR043.3-00.2 & \ang{ 287.75 } &  \ang{ 9.09} & HESS J1911+090 \cite{2010tsra.confE.201B, W49B} & no & --- \\ 
SNR045.7-00.4 & \ang{ 288.94} &  \ang{ 11.08} &  --- & no & ---  \\ 
SNR049.2-00.7 & \ang{ 290.81} &  \ang{ 14.14} &  W51 C \cite{W51} & no & yes \\ 
SNR074.0-08.5 & \ang{ 312.77} &  \ang{ 30.90} &  --- & yes & no \\ 
SNR078.2+02.1 & \ang{ 305.26} &  \ang{ 40.41} &  $\gamma$Cygni \cite{Aliu_2013} & no* & yes \\ 
SNR089.0+04.7 & \ang{ 311.15} &  \ang{ 50.42} & ---  & yes & no \\ 
SNR109.1-01.0 & \ang{ 345.41} &  \ang{ 58.83} & ---  & yes & no \\ 
SNR111.7-02.1 & \ang{ 350.85} &  \ang{ 58.83} & Cassiopeia A \cite{10.1093/mnras/stx2079}  & yes & no \\ 
SNR180.0-01.7 & \ang{ 84.55} &  \ang{ 27.86} & ---  & yes &  no\\ 
SNR189.1+03.0 & \ang{ 94.28} &  \ang{ 22.57} & IC 443 \cite{Albert_2007, Acciari_2009} &  yes & yes \\ 
SNR205.5+00.5 & \ang{ 98.91} &  \ang{ 5.87} & ---  & no* & no \\ 
\bottomrule
\end{tabular*}
\end{table}

Of the thirty GeV-detected SNRs in \cite{Acero_2016}, fifteen are contained within HAWC's field of view (declination between \ang{-26} and \ang{64}. Six of these are in source-rich regions of the galactic plane, dominated by emission from other objects. These six sources were excluded from this study. We then searched for TeV gamma-ray emission from the remaining nine sources, using the morphologies and spectral shapes from \cite{Acero_2016} as a baseline. 

For sources detected with a significance of more than $5\sigma$, the TeV spectrum was fit to the 1038-day dataset analyzed with the ground parameter energy estimator.
 
\section{Results}
\subsection{Dected SNRs}
Our of the nine SNRs in the sample, three were signficantly detected in HAWC data: SNR 049.2-00.7 (W 51), coincident with 2HWC J1922+140; SNR 078.2+02.1 ($\gamma$ Cygni), coincident with 2HWC J2020+403; and SNR 189.1+03.0 (IC 443). All were best fit with power-law spectra in the HAWC energy range, with no indication of curvature. The analysis of $\gamma$ Cygni requires a multi-source fit, and is treated in \cite{gCygniProceedings}.

\begin{figure*}[p]
\flushleft

\subfloat[]
	[SED of SNR049.2-00.7 (W51). MAGIC measurement from \cite{W51} shown in green. HAWC spectrum may include contamination from galactic diffuse emission or nearby sources.]
	{\includegraphics[width=0.48\textwidth]{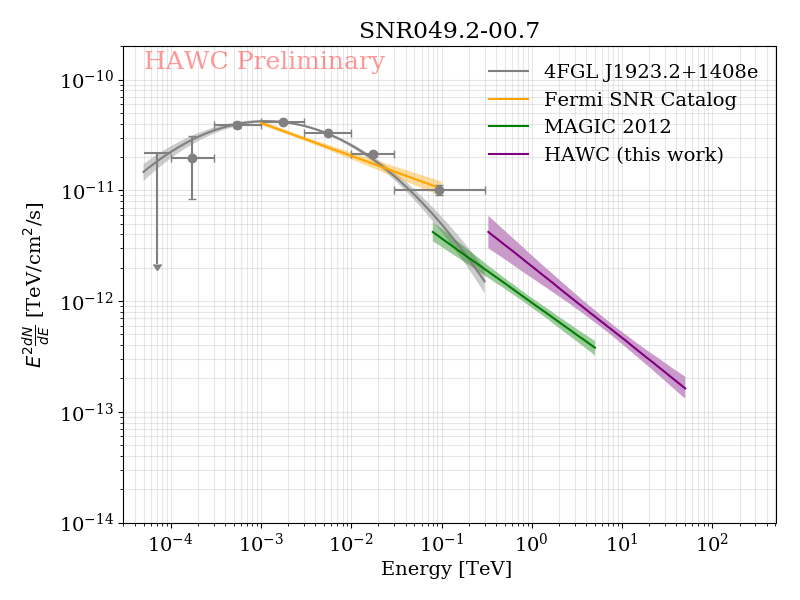}  }\hfill
\subfloat[]
	[SED of SNR109.1-01.0 (CTB 109)]
	{\includegraphics[width=0.48\textwidth]{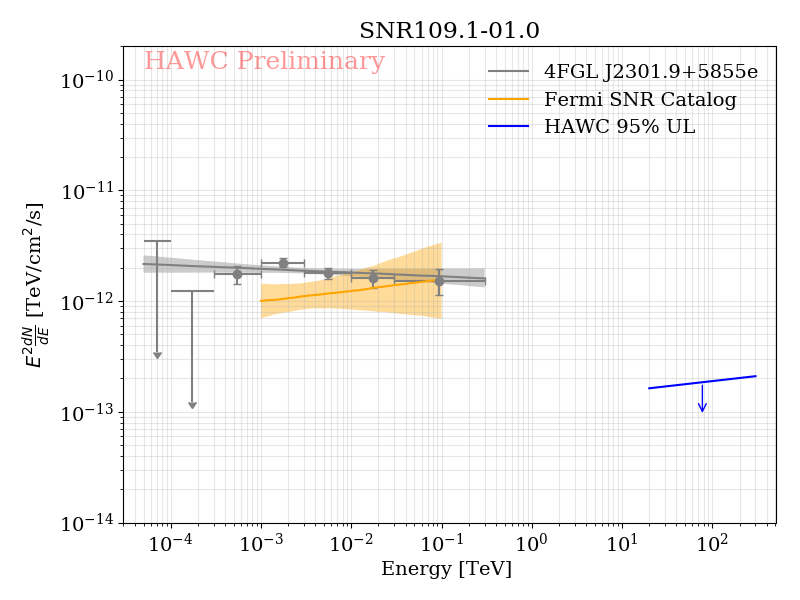}  }\\
\subfloat[]
	[SED of SNR 111.7-02.1 (Cassiopeia A). MAGIC measurement from \cite{10.1093/mnras/stx2079} in green.]
	{\includegraphics[width=0.48\textwidth]{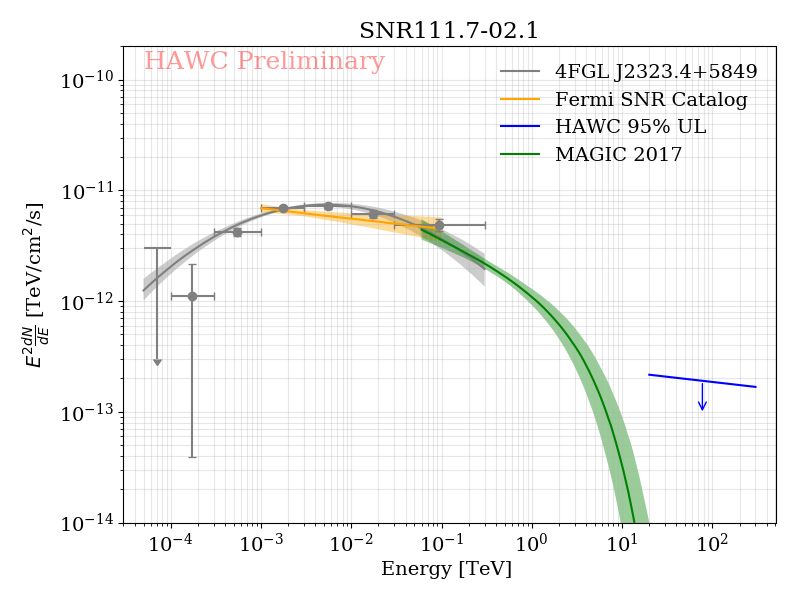}  }\hfill
\subfloat[SED of SNR180.0-01.7]
	[SED of SNR 180.0-01.7]
	{\includegraphics[width=0.48\textwidth]{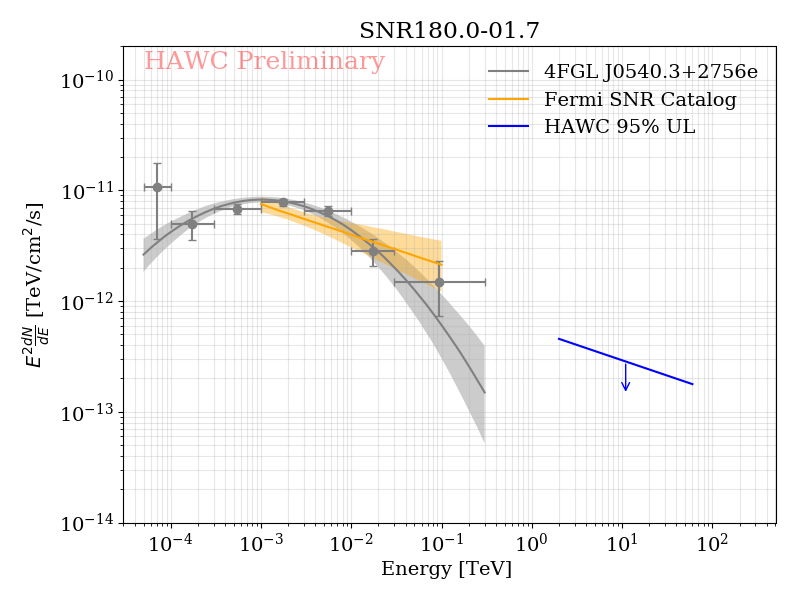}  }

\caption{Gamma-ray spectral energy densities (SEDs) of select GeV-detected SNRs. Plots show spectra from the 4FGL \cite{4FGL} in grey and from the Fermi-LAT SNR catalog \cite{Acero_2016} in orange. HAWC results (this study) are shown in purple (measurements) and blue (upper limits).}
\label{Fig:Spectra}
\end{figure*}

The best-fit energy spectrum for W 51 (SNR 049.2-00.7) is shown in \cref{Fig:Spectra} a). The HAWC measurement matches up well with the Fermi-LAT results, but with a somewhat softer index. The HAWC flux is a bit higher than the MAGIC measurement, which may be due to the presence of diffuse emission or other sources in the region. More detailed studies  of W51, as well as of IC 433, are ongoing and will be presented in future publications. 

\subsection{Non-detected SNRs}
Six SNRs in the sample did not show significant gamma-ray emission above the expectations from (hadronic) background. For these SNRs, 95\% CL upper limits on the integrated flux were calculated according to the prescription by Feldman \& Cousins \cite{Feldman:1997qc}, using the 1128-day dataset and the 9-bin analysis. \cref{tab:ULs} shows the results. For three objects, shown in \cref{Fig:Spectra} b)-d), the HAWC upper limits on the TeV gamma-ray fluxes are below the extrapolations from the GeV range, indicating a break or cut-off in the spectrum between HAWC and Fermi-LAT energies. Indeed, for SNR 111.7-02.1 (Cassiopeiea A), such a cut-off has been measured by the MAGIC experiment \cite{10.1093/mnras/stx2079}. The HAWC limits are fully consistent with the MAGIC spectrum. For SNR 180.0-01.7, no TeV counterpart has been identified, but the 4FGL results, derived from a larger energy range, confirm that the spectrum is curved. The HAWC limits are compatible with the 4FGL measurements.

\subsubsection{The Case of SNR 109.1-01.0}
SNR 109.1-01.0 (CTB 109) is a middle-aged SNR that has been detected and resolved in radio \cite{1981ApJ...246L.127H} and X-ray \cite{2004ApJ...617..322S, Sasaki:2013ldr}. Its age has been estimated to be between 9000 and 17000 years \cite{1981ApJ...246L.127H, 2004ApJ...617..322S}. CTB 109 shows an interesting semi-circular morphology, thought to be due to interactions of the remnant with sourrounding molecular clourd \cite{Sasaki:2013ldr}.

Both 4FGL and the 1FSC results show hard spectra (index around 2) without indications for spectral curvature in the GeV regime. Yet, no VHE counterpart has been identified, and the HAWC upper limits are more than an order of magnitude below the extrapolations from the FSC spectrum. This indicates that there must be a break or cut-off in the spectrum, similar to the much younger remnant Cassiopeia A.

In the 4FGL, CTB was reported as extended, with a radius of \ang{0.249}, and with a slighly softer spectral index (-2.03) compared to the 1FSC. We also looked for emission using the 4FGL morphology and spectrum; no significant emission was detected and the upper limits are still significantly below the extrapolation from the 4FGL.

\subsubsection{The Case of SNR 205.5+00.5}
SNR 205.5+00.5 shows the largest positive excess out of the non-detected remnants, corresponding to a significance of more than $3\sigma$. However, most of the emission in the region can be attributed to HAWC J0635+070 \cite{2018ATel12013....1B}, a TeV Halo candidate. With its (gaussian) extent of \ang[separate-uncertainty=true]{0.65\pm 0.18}, HAWC J0635+070 is considerably smaller than SNR 205.5+00.5 (disk radius of \ang{2.28 \pm 0.08}). The centroid of the HAWC source is also significantly offset (by \ang{1.1}) from the center of the GeV emission. However, to be conservative, the upper limits shown here have not been corrected for emission from HAWC J0635+070.

\section{Discussion and Conclusions}
We have searched for TeV gamma-ray emission from GeV detected SNRs with HAWC. Out of the nine SNRs in the sample that are not in source-confused regions, three SNRs were significantly detected with HAWC. Upper limits were derived for the remaining six objects. 

For the three detected SNRs, the energy spectra at TeV energies are best fit with power law shapes. The spectral indices are all softer than the GeV measured spectral indices, in line with previous measurements \cite{Acero_2016}. For three of the six non-detected SNRs, the HAWC upper limits are below the extrapolations from GeV measurements, again indicating a spectral cut-off or break between tens of GeV and tens of TeV.

Many supernova remnants are located in source-rich regions of the galactic plane, making it challenging to disentangle the TeV emission from (often extended) SNRs from neighgboring sources. Dedicated studies of the morphologies of these regions are needed to extract any detections or useful upper limits from HAWC data, similar to what has been done for $\gamma$ Cygni.

\begin{table}[p]
\caption{Results for non-detected SNRs. Table shows the nominal disk radius and spectral index from \cite{Acero_2016} used for the fit (sources without a radius were found to be point-like in \cite{Acero_2016} and accordingly, modeled as point sources for the HAWC analysis), the HAWC energy range (determined as in \cite{Abeysekara:2017mjj}, with the 75\%-TS-criterium), the upper limit on the gamma-ray flux integrated over the given energy range, and the TS for gamma-ray emission above background expectations.}
\label{tab:ULs}

\begin{tabular*}{\textwidth}{@{\extracolsep{\fill} } lcccccc}

\toprule
\textbf{SNR name} & \textbf{Radius} & \textbf{Index} & $E_{min}$ & $E_{max}$& \textbf{Flux UL} & \textbf{TS} \\
 & & & [TeV] & [TeV] & [cm$^{-2}$s$^{-1}$] & \\
\midrule
SNR074.0-08.5 &	\ang{1.74}	  	& 2.48	& 1.4 & 50  & \num[scientific-notation=true]{5.44E-13} & 0.11 \\
SNR089.0+04.7 &	\ang{0.97}		& 3.00	 	& 1.1 & 40	& \num[scientific-notation=true]{1.57E-12} & 0 \\
SNR109.1-01.0 &	--- 				& 1.91 	& 20  & 300 & \num[scientific-notation=true]{8.18E-15} & 0.05 \\
SNR111.7-02.1 &	---  			& 2.09	& 20  & 300 	& \num[scientific-notation=true]{9.37E-15} & 0.06 \\
SNR180.0-01.7 &	\ang{1.5} 		& 2.28 	& 2 	  & 60  & \num[scientific-notation=true]{1.77E-13} & -0.03 \\
SNR205.5+00.5 &	\ang{2.28} 		& 2.56 	& 1.7 & 40  & \num[scientific-notation=true]{1.48E-12} & 12.5 \\
\bottomrule 
\end{tabular*}
\end{table}

\clearpage

\setlength\bibitemsep{0.2\baselineskip}
\printbibliography

\end{document}